\documentclass[a4paper,11pt]{article}
\usepackage{pos}

\newlength{\bibitemsep}
\newlength{\bibparskip}
\let\oldthebibliography\thebibliography
\renewcommand\thebibliography[1]{%
  \oldthebibliography{#1}%
  \setlength{\parskip}{\bibitemsep}%
  \setlength{\itemsep}{\bibparskip}%
}

\usepackage[normalem]{ulem}

\title{Neutrino Target of Opportunity Sky Coverage and Scheduler for EUSO-SPB2
}
 \ShortTitle{Neutrino ToO Sky Coverage and Scheduler for EUSO-SPB2}

\author*[a]{Jonatan Posligua}
\author[b]{Tobias Heibges}
\author[b]{Hannah Wistrand}
\author[c,d]{Claire Gu\'epin}
\author[a]{Mary Hall Reno}
\author[e]{Tonia M. Venters}

\affiliation[a]{University of Iowa,Department of Physics and Astronomy,\\
  Iowa City, IA,  USA}

\affiliation[b]{Colorado School of Mines, Department of Physics,\\
Golden, CO, USA}

\affiliation[c]{University of Chicago, KICP,\\
Chicago, IL, USA}

\affiliation[d]{Laboratoire Univers et Particules de Montpellier,
\\
Montpellier, France}

\affiliation[e]{Goddard Space Flight Center,\\
Greenbelt, MD, USA}

\onbehalf{for the JEM-EUSO Collaboration}

\emailAdd{jonatan-posligua@uiowa.edu}

\abstract{Very-high-energy neutrinos can be observed by detecting air shower signals. Detection of transient target of opportunity (ToO) neutrino sources is part of a broader multimessenger program.  The Extreme Universe Space Observatory  on a Super Pressure Balloon 2 (EUSO-SPB2) Mission, launched on May 12, 2023, was equipped with an optical Cherenkov Telescope (CT) designed to detect up-going air showers sourced by Earth-skimming neutrinos that interact near the Earth’s limb. Presented here is an overview of the sky coverage and ToO scheduler software for EUSO-SPB2. By using the balloon trajectory coordinates and setting constraints on the positions of the Sun and Moon to ensure dark skies, we can determine if and when a source direction is slightly below the Earth’s limb. From a source catalog, CT scheduling and pointing is performed to optimize the search for high energy neutrinos coming from astrophysical sources. Some sample results for EUSO-SPB2 are shown.
}

\ConferenceLogo{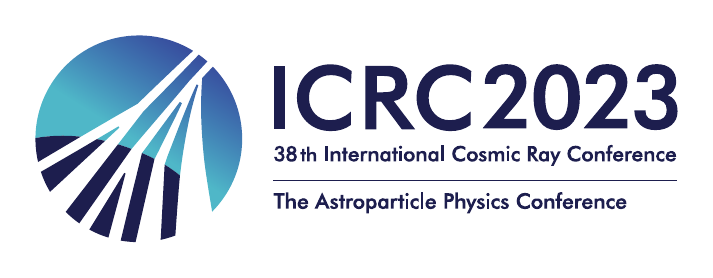}

\FullConference{%
The 38th International Cosmic Ray Conference (ICRC2023)\\
  26 July -- 3 August, 2023\\
  Nagoya, Japan}

\begin{document}
\maketitle

\section{Introduction}\label{sec:intro}

The Extreme Universe Space Observatory on a Super Pressure Balloon 2 (EUSO-SPB2) features an optical Cherenkov telescope (CT) and a fluorescence telescope (FT) flown on a NASA balloon. One of the goals of this mission was to search for very-high-energy (VHE) to ultra-high-energy (UHE) neutrino sources \cite{Eser:2021mbp}. Since VHE to UHE neutrinos (PeV-EeV) can be generated in distant environments such as active galactic nuclei (AGN), binary neutron star (BNS) mergers and binary black hole (BBH) mergers, these neutrinos may encode information about known and unknown fundamental physics happening in such environments \cite{Guepin:2022qpl}. 

The neutrino detection mechanism of EUSO-SPB2 consists of looking for up-going extensive air showers (EAS) generated by $\tau$-lepton decays. These EAS
produce Cherenkov radiation, which can be detected by the CT \cite{Venters:2019xwi,Cummings:2020ycz}. Ultra-relativistic muons in the Earth's atmosphere can also produce observable Cherenkov signals \cite{Cummings:2020ycz}. In order to have a transient neutrino source observable with the CT, several criteria must be met:
1) the source must be located behind the Earth, so that neutrinos coming from it can be converted into their corresponding charged leptons during their propagation through the Earth, 
2) the fraction of the Earth crossed by the charged leptons must be thin enough to allow them to escape and decay in the Earth's atmosphere,
3) only very little indirect light coming from the Sun or Moon reaches the telescope focal surface. 

The field of view (FOV) of the CT extends $\Delta\alpha= 6.4^\circ$ in altitude and $\Delta\phi=12.8^\circ$ in azimuth. Additional details describing the EUSO-SPB2 CT can be found in these proceedings \cite{Eser:2023icrc,Gazda:2023icrc}. To fulfil criteria 1 and 2, the pointing of the CT is such that the upper part of the field of view corresponds to the limb of Earth. Criterion 2 is significant because despite the fact that neutrinos must traverse a portion of the Earth to convert to charged leptons, over long distances, electromagnetic energy losses significantly reduce the energy of charged leptons so that either the energy of the air shower produced by its decay is reduced or it decays in the Earth before emerging \cite{Garg:2022ugd,Krizmanic:2023icrc}. Criterion 3 depends on the relative altitude and azimuth of the Sun and Moon with respect to the telescope, which change as a function of time. We discuss Criterion 3 in detail in Section 2 (Sun and Moon constraints) 

In the following, we describe a software developed in the context of the EUSO-SPB2 mission, which aims at producing schedules for EUSO-SPB2 CT observation of potential transient and steady sources of VHE neutrinos. Our main focus is follow-up observations of transient sources, such as blazar flares, gamma-ray bursts or binary neutron star mergers, hence we also refer to this observation mode as the Target of Opportunity (ToO) program.

\begin{figure}
    \centering
    \includegraphics[width=.75\textwidth]{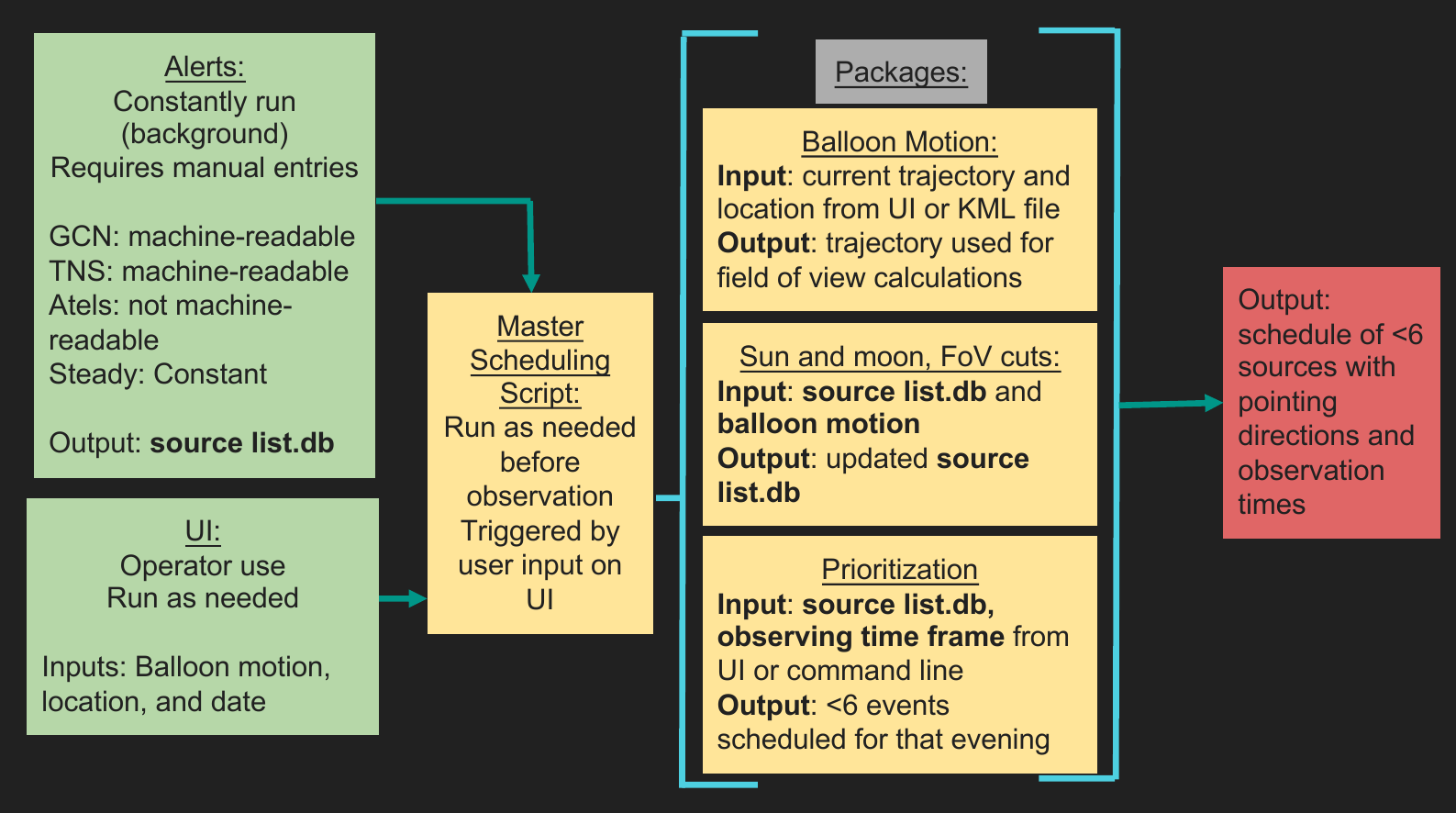}
    \caption{Flowchart of the software for the Target of Opportunity program.}
    \label{fig:flowchart}
\end{figure}

The aforementioned criteria are used to determine whether transient and steady sources can be observed during a given observing time window, and compute the times and pointing required for their observations. The balloon trajectory is taken into account, as its latitude and longitude impacts the list of observable sources. Prioritization criteria are used to select sources to be scheduled for observations. Moreover, we develop a strategy to observe extended sources such as TA old/new hotspots \cite{TelescopeArray:2014tsd,TelescopeArray:2022gei} or localization probabilities from gravitational wave signals \cite{LVKAlerts}. Further details on the ToO program and prioritizations of ToO neutrino sources appear in these proceedings \cite{Heibges:2023icrc,Wistrand:2023icrc}. A flowchart of the ToO software is shown in figure  \ref{fig:flowchart}.

\section{Structure of the ToO software}

The ToO software builds a list of observable sources for each observation period, and proposes an observation schedule, comprised of a sub-sample of these observable sources. From a transient and steady source database, observable sources are determined using criteria 1, 2 and 3 (described in section~\ref{sec:intro}). In the following we detail how we build the source database, how we account for the balloon motion, the impact of the Sun and the Moon on the observation window, observability constraints from the field of view, and the prioritization strategy. The user interface is also described.\vspace{0.1cm}

\textbf{Transient and steady source database:} The first module of the ToO software is a listening module designed to filter alerts from various alert networks, such as the Gamma-ray Coordination Network (GCN) and the Transient Name Server (TNS). GCN and TNS alerts are available in machine-readable formats \cite{Heibges:2023icrc}. Several criteria determine which alerts are added to a source database: we select alerts corresponding to potential sources of VHE neutrinos. The software updates the source database on a continuous basis. In addition, we incorporate Astronomer's Telegram (ATel) alerts. These are not machine readable, thus are hand-processed. A catalog of steady sources is also included, and additional sources can be added by the user.  With these various components, a combined source list is produced in a generic database format (db).\vspace{0.1cm}

\textbf{Balloon trajectory:} In order to determine the observing schedule, the software must account for the trajectory of EUSO-SPB2. During the mission, a KML file is provided by the Columbia Scientific Balloon Facility (CSBF) that provides the balloon's projected latitude, longitude, and altitude for the next three days in $6~{\rm h}$ bins. 
The nominal float altitude of the 18MCF super-pressure balloon used for the Wanaka flight is 33 km, with daily variations in float altitude much less than 1 km.
Moreover, super-pressure balloons taking off from Wanaka (New Zealand) are expected to reach high altitude winds circling around Antarctica. 

\begin{figure}
    \centering
    \includegraphics[width=0.46\textwidth]{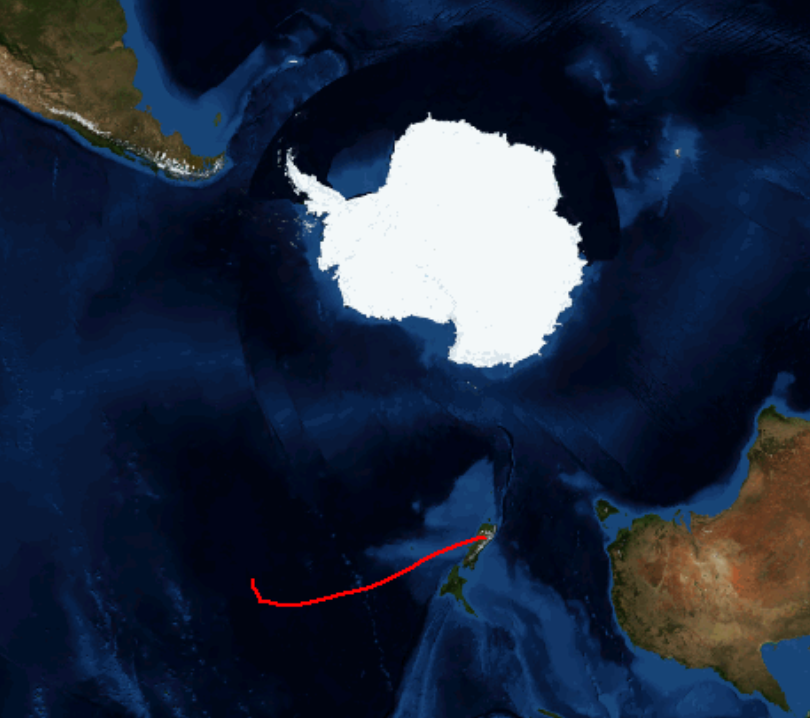}
    \includegraphics[width=0.44\textwidth]{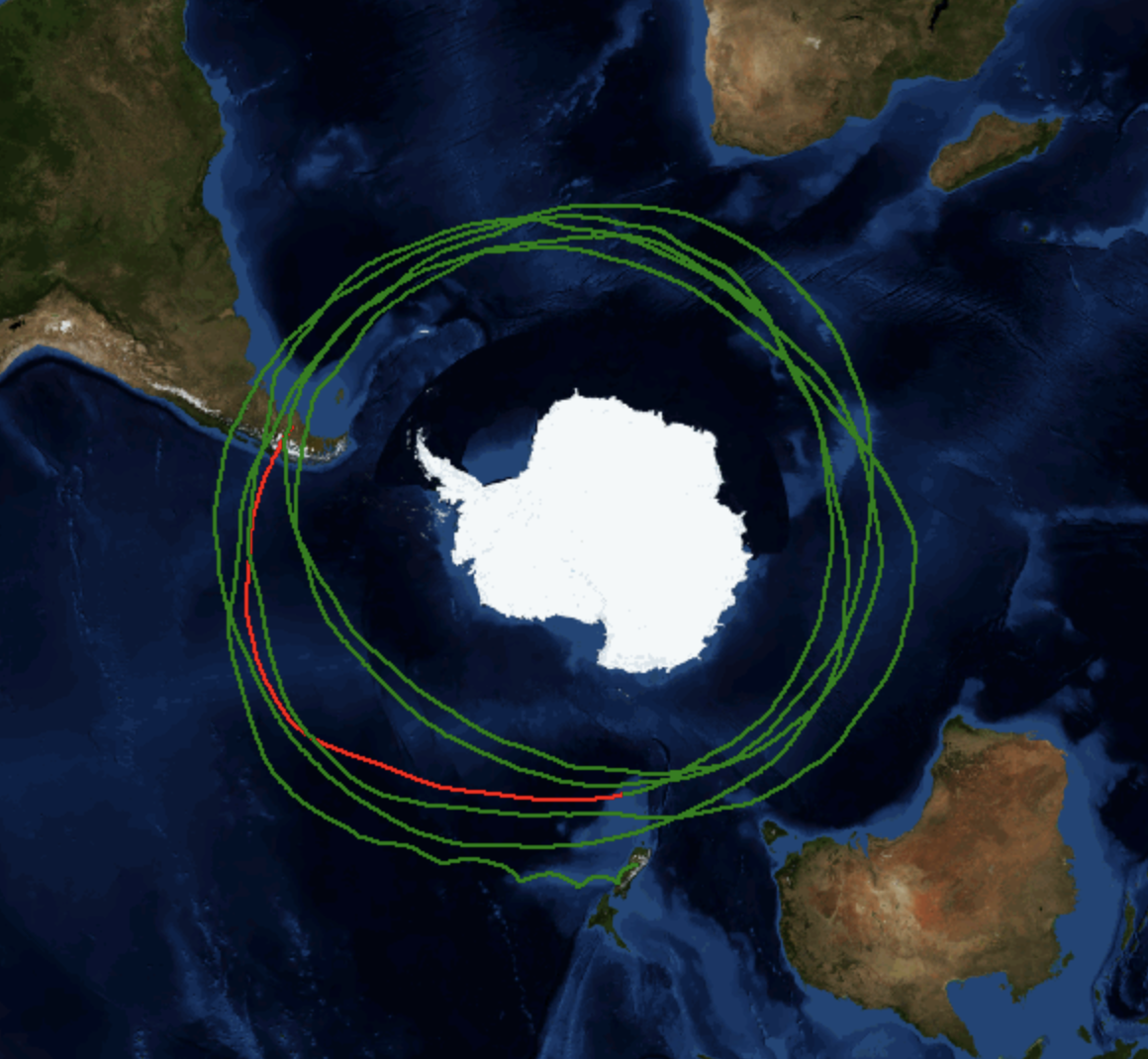}
    \caption{Payload trajectories of EUSO-SPB2 (left) and SuperBIT (right). Figures from  www.csbf.nasa.gov.}
    \label{fig:trajectories}
\end{figure}

EUSO-SPB2 was launched from Wanaka on May 13, 2023. Unfortunately, a bad leak in the balloon limited its flight to 1 day, 12 hours and 53 minutes, with termination in the Pacific Ocean. The SuperBIT super-pressure balloon was launched on April 15, 2023 from Wanaka, New Zealand, and its flight lasted for 39 days, 13 hours and 35 minutes. Both trajectories are illustrated in figure~\ref{fig:trajectories}. The software was extensively tested using the KML files of the SuperBIT flight. The tests mainly focused on computing the observing window and the field of view as a function of balloon location and time, and produce mock observing schedules. A sample observing schedule for the second night of the EUSO-SPB2 flight is shown below. The software described in this proceedings is designed to be applicable for any trajectory and time period, and will eventually be adapted to satellites and ground-based telescopes.\vspace{0.1cm}

\begin{figure}[ht]
    \centering
        \includegraphics[width=.55\textwidth]{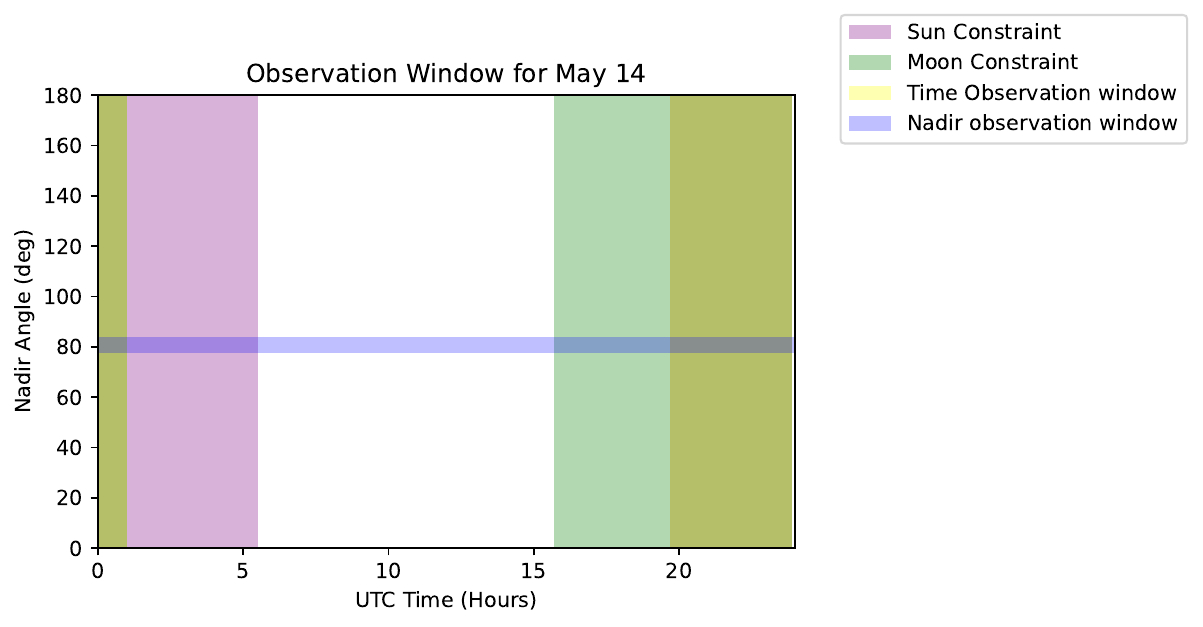}
    \includegraphics[width=0.38\textwidth]{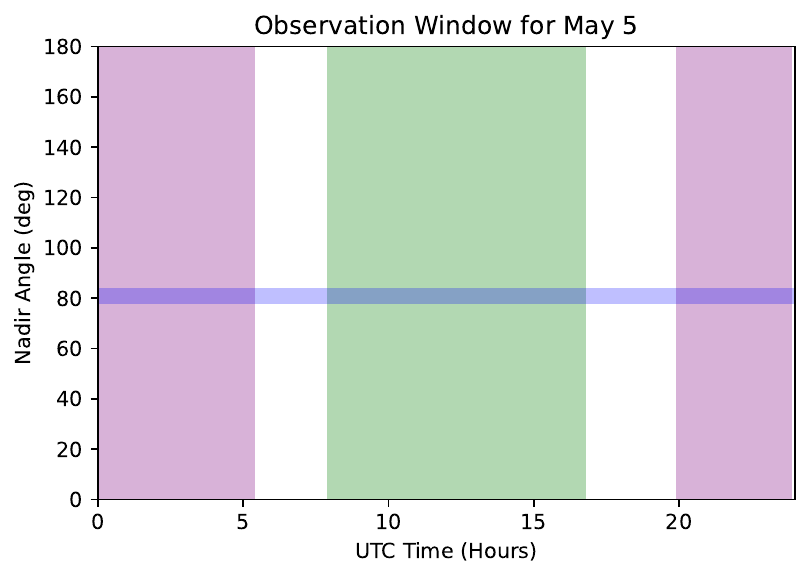}    
    \caption{Observations windows for two dates in May, 2023 for observations at Wanaka, NZ. The vertical shaded bands show times when the Sun and Moon altitude constraints are satisfied (i.e., the Sun and Moon have set). Both the Sun and Moon satisfy observing constraints only during the time periods when the green and purple bands overlap (the time observation window).}
    \label{fig:SunMoonWindows}
\end{figure}

\textbf{Sun and Moon constraints:} In order to observe any neutrino source with optical Cherenkov radiation, the sky must be dark. For any day and location for which we run the observation schedule, the observation window is open when the Sun and Moon have set. In this work, we consider those altitude thresholds to be -24$^\circ$ for the Sun (astronomical night) and -6$^\circ$ for the Moon (below the limb of the Earth) unless the Moon's illumination is small, below $0.05$. Figure~\ref{fig:SunMoonWindows} shows the observation window for two different days. 
The purple shaded region shows the time period when the Sun is set. The green shaded region shows times when the Moon is set (when the illumination is larger than 0.05). The observation window is represented by the intersection between the vertically-spanned purple and green regions. In this example, sources can be observed on May 14, but not on May 5 because the Moon and Sun altitude criteria are not simultaneously satisfied.\vspace{0.1cm}

\textbf{Field of view constraints:}
In the VHE neutrino observation mode, the CT is pointed such that the upper part of the field of view corresponds to the Earth's limb. It extends $\Delta\alpha= 6.4^\circ$ in altitude and $\Delta\phi=12.8^\circ$ in azimuth. Therefore, the field of view constraint for observability is simply given by a geometric constraint: during the observing window, there is a time period for which the altitude of the source is contained between $-6^\circ$ (Earth's limb) and $-12.4^\circ$ (lower end of the FOV). This geometric constraint does not include any constraint in azimuth as we can rotate the CT to point to any location in azimuth. This geometric constraint is represented by the blue shaded region in figure~\ref{fig:SunMoonWindows}. Observable ToO neutrino sources must pass through the intersection of the purple, green and blue bands. For all sources satisfying the FOV constraint, the Sun/Moon/FOV module computes the range of times and azimuths for which these sources are observable. These quantities are used to compute the observation schedule, and inform when and where to point the telescope to observe the sources.\vspace{0.1cm}

\textbf{Prioritization}: The source catalog \cite{Wistrand:2023icrc} 
contains an extensive number of sources that fulfil the observability criteria during an observation run. To compute the observation schedule, prioritization is required, because the observation time is limited, and only a few  re-pointings (4 to 5) can be done during each observation run. Therefore, we favor sources that have the highest potential to emit detectable fluxes of VHE neutrinos \cite{Wistrand:2023icrc}. 
From first to last in level of priority, the type of sources are ranked as: 1) Galactic supernovae, 2) binary neutron stars (BNS) and black hole neutron star (BH-NS) mergers, 3) nearby tidal disruption events (TDE), 4) flaring blazars or active galactic nuclei (AGN), 5) gamma-ray bursts (GRB), 6) supernovae, 7) other transients and 8) steady sources outside of the IceCube sensitivity region. The scheduler accounts for this priority ranking in setting observation times and telescope azimuthal pointing directions for a given night. In the prioritization algorithm, we also account for the maximum time during which sources can cross the field of view of the detector. These times can vary between 20~mins and 1~h~20~mins. In each ranking category, we favor the sources that can be observed the longest, considering that observation time is a good proxy for the acceptance. This assumption will be validated by detailed simulation computing the acceptance as a function of energy and nadir angle.  \vspace{0.1cm}

\textbf{User interface:} In order to create a software accessible for a variety of users, we developed a user-friendly interface. The software modules used to collect alerts, generate source catalogs, and generate the observation schedules, are incorporated in a GitLab distribution. The user interface allows users to choose sets of initial parameters and easily run the various modules, and provides human-readable results for observation parameters of interest each time the scheduling software is run.

\section{Scheduling observations}

\begin{figure}
    \centering    \includegraphics[width=0.73\textwidth]{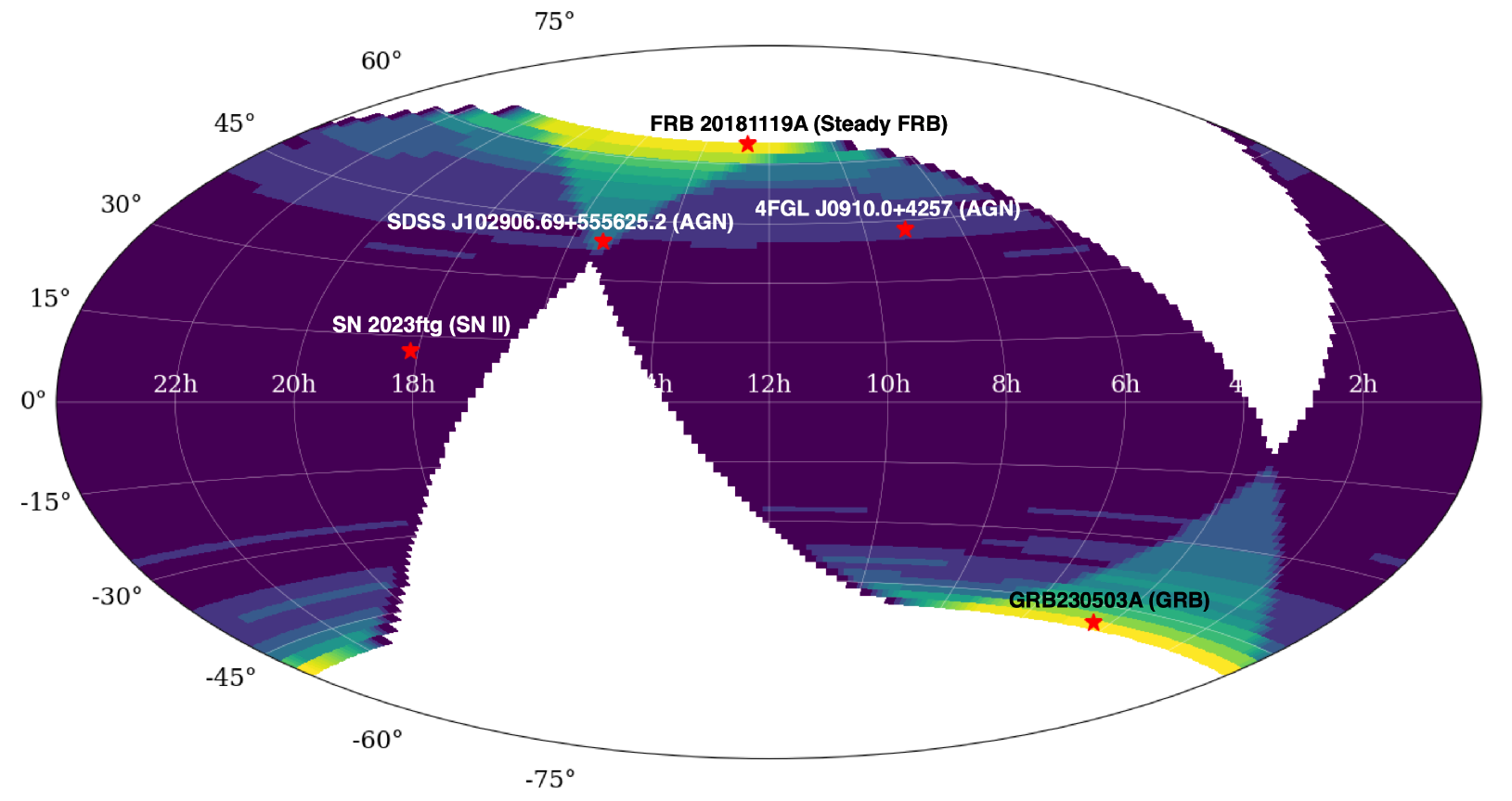}
   \caption{Sky plot of scheduled sources in equatorial coordinates (RA,DEC), for May 14, 2023 considering the projected EUSO-SPB2 balloon path. The colored regions represent the maximum exposure as a function of (RA,DEC), with lighter colors representing higher exposures.}
    \label{fig:pointsourceviz_may14}
\end{figure}

\begin{figure}
    \centering
     \includegraphics[width=.49\textwidth]{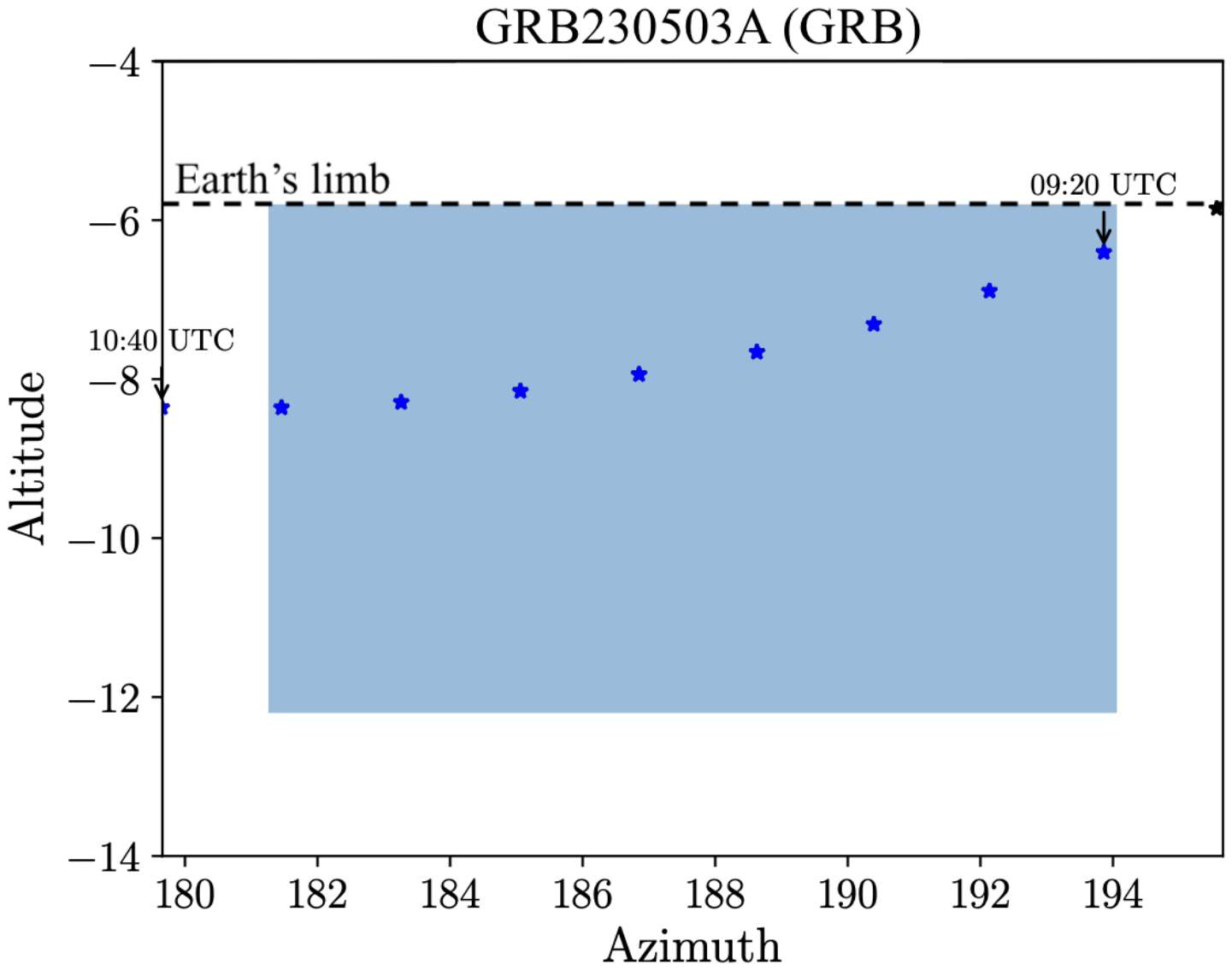}
     \includegraphics[width=.49\textwidth]{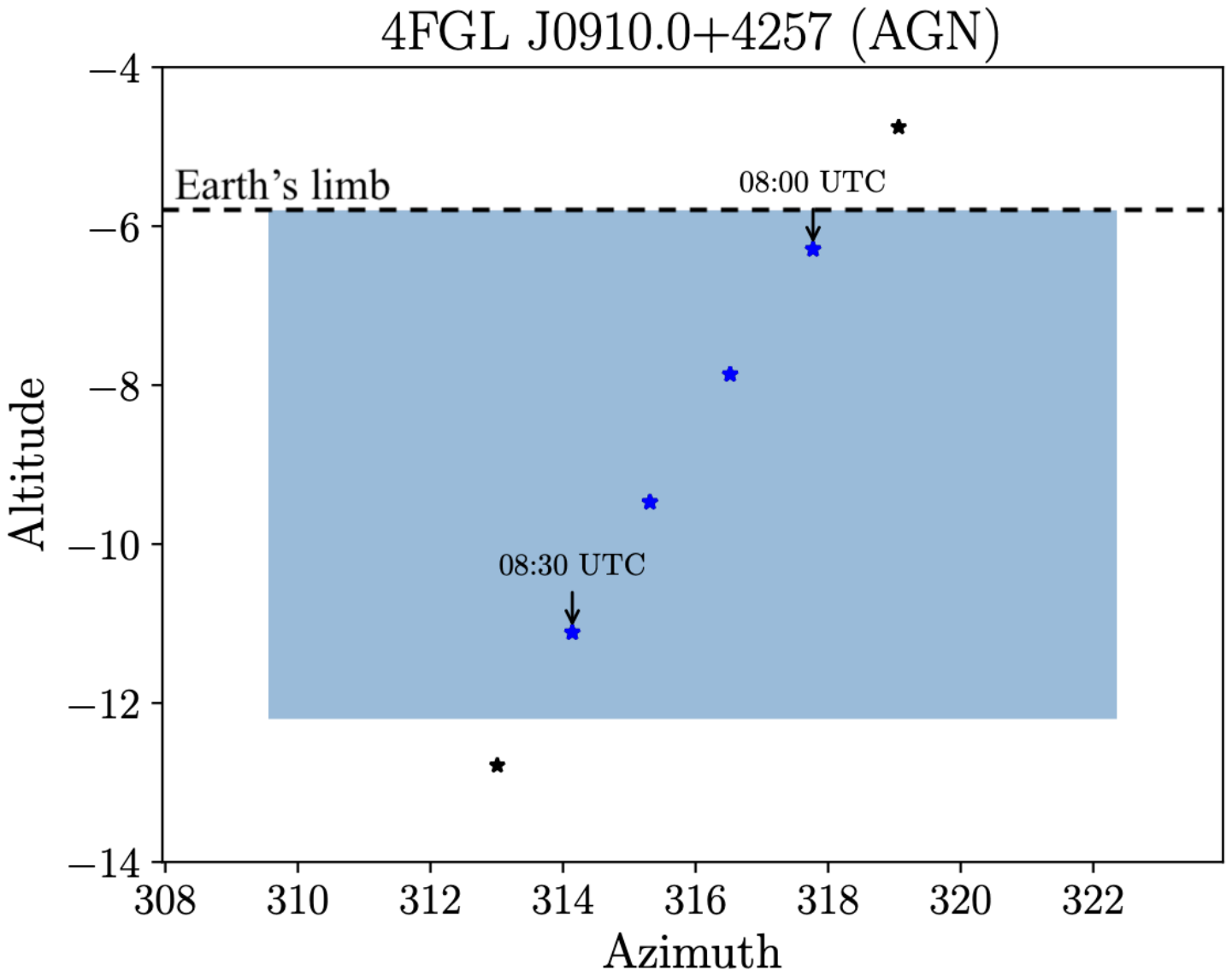}
   \caption{Two of the scheduled sources for May 14, 2023 passing through the CT FOV (blue shaded region), as a function of the altitude and the azimuth. Blue stars represent the expected position of the source, in 10~mins bins. Reference times label two source locations in the FOV. Source enter the FOV at the larger azimuth.}
    \label{fig:pointsourcetraj}
\end{figure}

\textbf{Point sources:}
An example for point source scheduling has been computed for May 14, 2023 \cite{Wistrand:2023icrc}, using the projected trajectory of EUSO-SPB2. The observing window spans from 5:10am (UTC) to 10:45am (UTC). 
The location of these 5 sources is illustrated in a sky map in equatorial coordinates in figure~\ref{fig:pointsourceviz_may14}. Following the prioritization strategy, the schedule was first filled with sources corresponding to higher levels of priorities, such as GRBs and AGN. When several sources are observable in a priority tier, the source selected is the one that crosses the field of view for the longest time, as we expect time to be a good proxy for acceptance. Two source trajectories are shown in figure~\ref{fig:pointsourcetraj}, when they cross the CT field of view. These trajectories illustrate the important difference in observing time between sources that skim the Earth's limb (in our example the GRB) and the others. More details about the scheduling and a specific schedule for May 14, 2023 appears in ref. \cite{Wistrand:2023icrc}.

\textbf{Extended sources:} For point sources, each source has a well-defined set of sky-coordinates. However, some of the sources that are candidates for neutrino production are extended regions in the sky, e.g., localization regions determined from gravitational wave detection \cite{LVKAlerts} and the TA hotspot \cite{TelescopeArray:2014tsd,TelescopeArray:2022gei}. For these types of sources, we only have knowledge of the probability of localization at each point contained within a given region. 
Our goal is to optimize pointing of the telescope based on the duration of observations and the probability of localization for extended sources. 
 
\begin{figure}[htp]
    \centering
    \includegraphics[width=.5\textwidth]{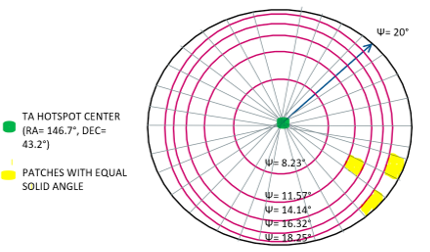}
    \caption{Equal solid angle patches delineated by angle $\psi$ (red) curves and azimuth (blue) lines.}
    \label{fig:geometry}
\end{figure}

The approach we take consists of dividing up the extended source into equal-sized solid angle patches, as shown in figure \ref{fig:geometry}. These patches are treated as effective point sources located in the middle of each patch. With the sky coordinates and localization probability of each patch, the extended source module evaluates the observation parameters for each of these effective point sources: initial/final time and azimuth, total observation time, total span in azimuth.
Then, we determine the pointing in azimuth for the CT that yields the maximum expectation value for the total observation time. This total time refers to the probability-weighed time summed over all effective point sources that manage to come into the CT FOV during an observation run.

An example of the application of this procedure is to (RA,DEC)=(146.7$^\circ$,43.2$^\circ$), the old TA hotspot location, where we use a Gaussian probability distribution $f(\psi,A,\sigma)=A\exp(-\psi^2/(2\sigma^2)$) for the probability of neutrinos coming from an angle $\psi$ from the center of the hotspot, with $A=0.07746$ and $\sigma=10.3^\circ$
\cite{TelescopeArray:2014tsd}. For the computation, we consider the date May 11, 2023 and the balloon location in Wanaka, NZ. Dividing the hotspot in 150 patches, the patch with the highest product of probability and observing time is obtained for a CT pointing to the azimuth $357.55^\circ$. On the same date, optimizing instead the expectation value yields a telescope azimuth pointing to $348^\circ$. Typically, optimizing the expectation values of this extended source gives an azimuth pointing between $10^\circ-30^\circ$ away from the best single patch in the TA hotspot. The same procedure can be applied for sample localization probabilities from Ligo/Kagra/Virgo.

\section{Conclusion and prospects}

The software described here was developed to support the EUSO-SPB2 science goals for neutrino ToO observations. It is designed to produce schedules for CT operators to use as a guide each night, as they adapt to cloud conditions and schedule other science observing goals.

The software described is modular and can be used for various missions aiming at observing VHE neutrinos, with and without pointing. It will be used for future JEM-EUSO balloon missions. Moreover, it will be applied to the trajectory of Mini-EUSO, a JEM-EUSO instrument equipped with a mini-fluorescence telescope installed on the International Space Station. The software could also be applied to Terzina, a project for a CT telescope on a small satellite \cite{Aloisio:2023icrc}. The modules that produce catalogs and identify which sources pass through the FOV could be used by PUEO, a radio detection mission on a balloon, and by ground-based instruments such as GRAND. After the software is finalized and documented, it will be open source and available via GitLab.\vspace{0.1cm}

\noindent{\bf Acknowledgements} -- The authors acknowledge support by
NASA awards 11-APRA-0058, 16-APROBES16-0023, 17-APRA17-0066, and NNX17AJ82G, NNX13AH54G, 80NSSC18K0246, 80NSSC18K0473, 80NSSC19K0626, and 80NSSC18K0464, the French space agency CNES, National Science Centre in Poland grant n. 2017/27/B/ST9/02162, and by ASI-INFN agreement n. 2021-8-HH.0 and its amendments.  This research used resources of the National Energy Research Scientific Computing Center (NERSC), a U.S. Department of Energy Office of Science User Facility operated under Contract No. DE-AC02-05CH11231. We acknowledge the NASA Balloon Program Office and the Columbia Scientific Balloon Facility and staff for extensive support. We
also acknowledge the invaluable contributions of the administrative and technical staffs at our
home institutions.

%\bibliographystyle{JHEP}
%\bibliography{references}

\providecommand{\href}[2]{#2}\begingroup\raggedright\endgroup

\clearpage

\newpage
{\Large\bf Full Authors list: The JEM-EUSO Collaboration\\}
%{\scriptsize (author-list as of July 15th, 2023 with reorganized affiliations)} \hspace{0.6cm}
%{\scriptsize (version  \today{} \currenttime{})}
%\vspace*{0.5cm}

\begin{sloppypar}
{\small \noindent
S.~Abe$^{ff}$, 
J.H.~Adams Jr.$^{ld}$, 
D.~Allard$^{cb}$,
P.~Alldredge$^{ld}$,
R.~Aloisio$^{ep}$,
L.~Anchordoqui$^{le}$,
A.~Anzalone$^{ed,eh}$, 
E.~Arnone$^{ek,el}$,
M.~Bagheri$^{lh}$,
B.~Baret$^{cb}$
% ...
}
\end{sloppypar}
\vspace*{.3cm}

%%\newpage
{ \footnotesize
\noindent
% Algeria - 2 institutes
% ...
$^{cb}$ Universit\'e de Paris, CNRS, AstroParticule et Cosmologie, F-75013 Paris, France\\
% ...
$^{ed}$ Istituto Nazionale di Fisica Nucleare - Sezione di Catania, Italy\\
% ...
$^{eh}$ INAF - Istituto di Astrofisica Spaziale e Fisica Cosmica di Palermo, Italy\\
% ...
$^{ek}$ Istituto Nazionale di Fisica Nucleare - Sezione di Torino, Italy\\
$^{el}$ Dipartimento di Fisica, Universit\`a di Torino, Italy\\
% ...
$^{ep}$ Gran Sasso Science Institute, L'Aquila, Italy\\
% Japan - 7 institutes 
% ...
$^{ff}$ Nihon University Chiyoda, Tokyo, Japan\\ 
% ...
$^{ld}$ University of Alabama in Huntsville, Huntsville, AL, USA\\
$^{le}$ Lehman College, City University of New York (CUNY), NY, USA\\
% ...
$^{lh}$ Georgia Institute of Technology, USA\\
% ...
}

\end{document}